\documentclass{camera}
\usepackage{graphicx}  % uncomment this if your want to insert figures

\begin{document}

%%%%%%%%%%%%%%%%%%%%%%%%%%%%%%%%%%%%%%%%%%%%%%%%%%%%%%%%
% The title, only the first letter capitalized; if you want to split it in
% two or more lines, put a \\ macro at each line break
% example: 
%   \title{Title: first line\\ second line}
%
\title{Comparison of fragment partitions produced in peripheral and central collisions.}

%%%%%%%%%%%%%%%%%%%%%%%%%%%%%%%%%%%%%%%%%%%%%%%%%%%%%%%%
% The author(s), separated by commas; do not put a
% comma before the last author, use instead the \and
% macro which produces a normal ``and'' in the
% caps/small caps context
%
\author{E. Bonnet$^{a,b}$, B. Borderie$^a$, N. Le Neindre$^a$ \and
M.F. Rivet$^a$\\ for the INDRA and ALADIN collaborations.}
%%%%%%%%%%%%%%%%%%%%%%%%%%%%%%%%%%%%%%%%%%%%%%%%%%%%%%%%
\organization{a) IPN Orsay; b) GANIL}
\maketitle
\begin{abstract}
Ensembles of single-source events, produced in peripheral and central collisions and correponding respectively to 
quasi-projectile and quasi-fusion sources, are analyzed. After selections on fragment kinematic properties, 
excitation energies of the sources are derived using the calorimetric method and the mean behaviour of
fragments of the two ensembles are compared. 
Differences observed in their partitions, especially the charge asymmetry,
can be related to collective energy deposited in the systems during the collisions.
\end{abstract}
%%%%%%%%%%%%%%%%%%%%%%%%%%%%%%%%%%%%%%%%%%%%%%%%%%%%%%%%
% Write the text starting from here and using the usual
% LaTeX commands.
%
\section*{Introduction}
\indent
Heavy-ion collisions at intermediate energies give access with good
detection efficiency to two kind of multifragmenting sources: Quasi-Fusion (QF) sources produced
in central collisions and Quasi-Projectile (QP) sources produced
in peripheral collisions. Open questions are: how do the hot sources produced explore
the phase diagram after their formation? How the trajectories in this phase diagram influence
the de-excitation properties? For central collisions, experimental
results \cite{expa} show that radial collective energy is present in addition to thermal pressure during the
multifragmentation process. This presence of radial collective energy is the consequence of the compression/expansion
cycle where the hot system passes through before its multifragmentation. In
the following, we shall present
briefly experimental data and event selections. In a second part, a method to extract expansion energy in QP
sources will be presented and their trajectories in the phase diagram will be deduced. The third part will be
devoted to a comparison of fragment partitions from QP and QF sources.
Finally we shall conclude and draw some outlooks.

\section{Selection of event ensembles.}
Data collected by the multidetector INDRA are used. For central collisions, we chose the Xe+Sn reaction at five
bombarding energies (25, 32, 39, 45 and 50 MeV/A) performed at GANIL and for peripheral collisions
the Au+Au reaction at 80 MeV/A performed at GSI. These two systems are close
in size (ratio of about 1.3) and the chosen bombarding energies give a
sufficient overlap in excitation energy. For both central and peripheral
collisions, we only keep correctly detected and measured events which have a total detected charge greater
than 80\% of the system charge. Well characterized events are obtained for QF sources using flow angle
selection \cite{tflo} and we propose a new method of selection for QP
sources \cite{these}. In both cases fragments are defined as products with charge greater or equal to 5.

\subsection{QF sources selection}
The kinetic energy tensor is computed event by event in the centre of mass of the reaction. This global
variable, provides information such as the flow angle ($\theta_{flow}$) which
characterizes the event main axis with
respect to the beam direction. Events with $\theta_{flow}>60^{o}$ are kept for the five bombarding energies.

\subsection{QP sources selection}
A new selection is proposed \cite{these} to select events with all fragments associated
to the de-excitation of QP sources. The goal of that selection is to remove
events which contain emissions from the participant
zone which populate the same velocity space region than QP sources; indeed
such emissions can blur the QP de-excitation 
pictures. The selection is based on a compactness criterion in the velocity space of fragments. The dedicated
variable called VarDyn combines two observables (eq.\ref{vardyn}). The first
one is the reconstructed velocity
of the sources ($\beta_{qp}$) in the centre of mass of the reaction. It provides information on the dissipation
of collisions. More dissipative is the collision lower is the 
$\beta_{qp}$ value, which tends asymptotically to zero for central collisions. The second observable is
the mean relative velocity between fragments ($\beta_{rel}$) which is a
measure of the compactness of QP events in velocity space.
If one or several fragments come from the mid-rapidity (MR) region, we
expect larger values than in the case
where all secondary fragments are localized around the reconstructed velocity sources. The VarDyn observable is
defined as the ratio between these two observables. Well defined QP sources are selected asking for the
compactness criterion : VarDyn=$\beta_{qp}$/$\beta_{rel}$$>$1.5. Events with fragments in the MR region
and the most dissipative collisions are thus rejected.

\begin{eqnarray}
\beta_{rel} = \frac{2}{M_{frag}(M_{frag-1})} \sum_{i<j}^{M_{frag}}|\vec{\beta_{i}}-\vec{\beta_{j}}| \qquad \beta_{qp} = \big|\sum_{i}^{M_{frag}}\frac{\vec{p_{i}}}{E_{i}}\big| \label{vardyn} \\ 
\beta^{(N)}_{rel}=\frac{\beta_{rel}}{\sqrt{<Z>(Z_{s}-<Z>)}} \label{norm}
\end{eqnarray}

\subsection{Choice of excitation energy as control parameter.}
To compare the two sets of events corresponding to QP and QF sources, we use the excitation energy ($E^{*}$)
obtained with a calorimetric procedure. This procedure consists in an event by event total energy balance.
The same algorithm and the same hypotheses on parameters are used for both QP and QF sources \cite{these}.
Then we deduce
excitation energies and sizes of the reconstructed sources ($Z_{s}$). A common $E^{*}$ range between 4 and
12 MeV/A is populated by the two types of sources for which a size ratio of
about 1.2-1.3 is deduced.

\section{Radial collective energy in peripheral\\collisions -
comparison to central collisions.}
\begin{figure}
\begin{center}
\includegraphics[scale=0.45]{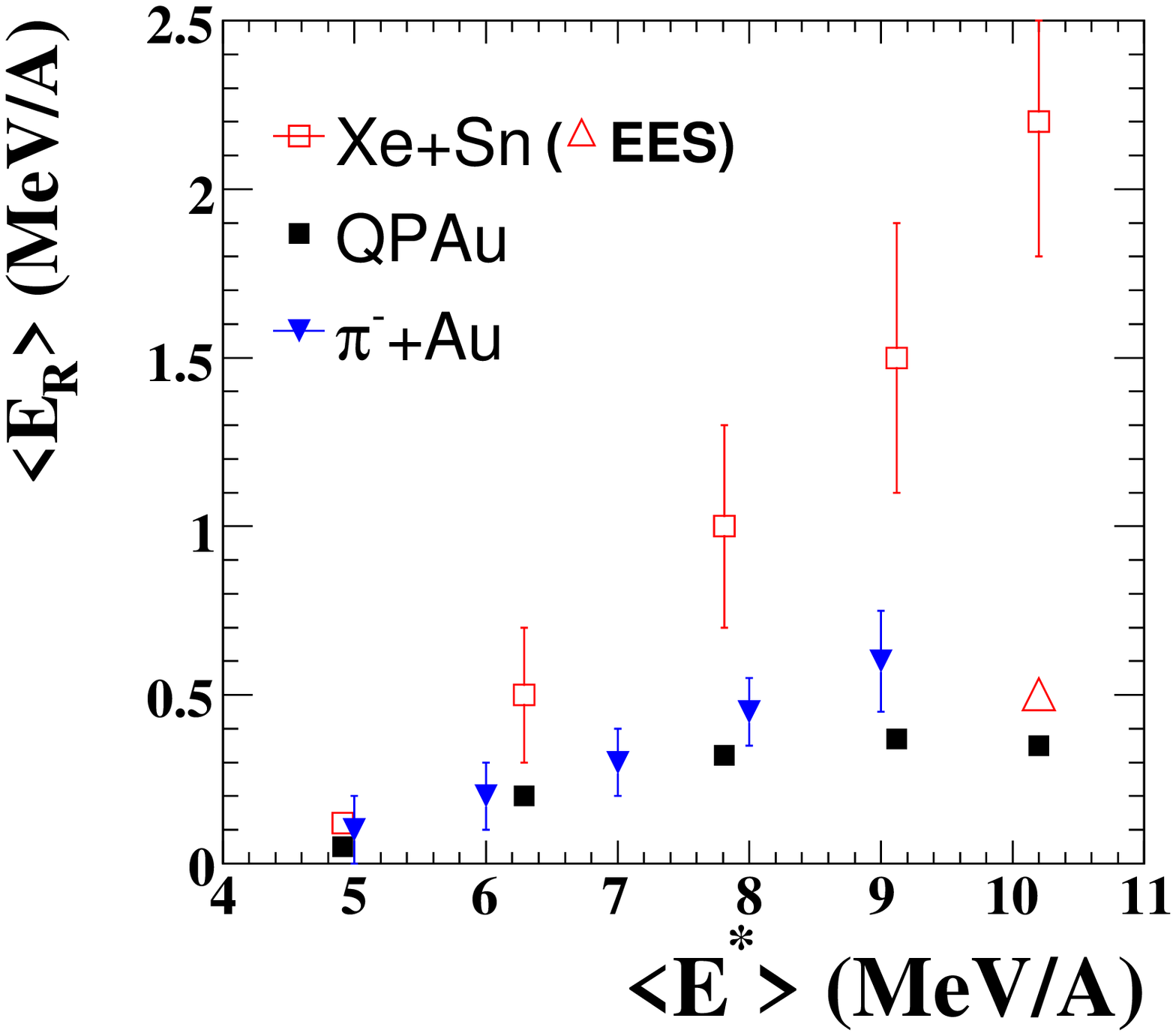}\caption{}
%\caption{Radial collective energy as a function of source excitation energy.
%See text for the different symbols}
\label{fig01} % optional figure label, must be unique
\end{center}
\end{figure}
To obtain experimental information on collective radial energy in QP
sources, we use the $\beta_{rel}$ observable (previously
defined in eq.\ref{vardyn}) and calibrate it using QF data from central collisions.
However to take into account Coulomb and size effects
we normalize $\beta_{rel}$ with a term which takes into account both effects through
the mean Coulomb energy
seen by the average fragment charge of each event due to the other ones.
 As we are dealing with velocities we take the root
square of this factor and obtain the normalized relative velocity
$\beta^{(N)}_{rel}$ (eq.\ref{norm}).
At $E^{*}\sim$5 MeV/A, which corresponds to the QF event ensemble of Xe+Sn
at 25 MeV/A incident energy, normalized relative velocities are similar for QF and QP sources.
When $E^{*}$ increases, the increase of $\beta^{(N)}_{rel}$ is stronger for QF sources. At $E^{*}\sim$10 MeV/A,
which corresponds to QF events from Xe+Sn at 50 MeV/A, the mean value of the normalized
relative velocity for QF sources is twice the value of QP ones.\\
To link this observation with the amount of collective radial energy in
sources, we have made a calibration
of $\beta^{(N)}_{rel}$ using four published values of collective radial energy for central collisions \cite{expa}.
These values are obtained for the same Xe+Sn reactions from 32 to 50 MeV/A
using a statistical model (SMM)
where the collective radial energy is a free parameter and can be tuned to reproduce experimental kinetic energies.
This calibration allows us to obtain an estimate of the collective energy
for Xe+Sn at 25 MeV/A and for the five corresponding
excitation energy values for QP sources.
Figure \ref{fig01} summarizes all the values of radial collective energies as
a function of excitation energy. We have also added other experimental
values obtained in the study of the multifragmentation of gold induced by
hadrons ($\pi^{-}$) by the ISIS
collaboration \cite{isis}. In this type of
reactions, a collective radial energy only due to thermal pressure is expected.
 For central collisions (open squares), we observe a strong correlation
between excitation energy
and radial collective energy whereas for QP sources (full squares), the trend is close to
ISIS values, which indicates that the main part of the
radial collective energy is due to thermal pressure. For the three types of
reactions, we see that the onset of collective radial energy is localized around $E^{*}$=5 MeV/A.
Concerning the contribution to
radial collective energy coming from thermal pressure for QF sources, a
calculation with the EES model \cite{ees} for a source at around $E^{*}$=10 MeV/A
gives a value of 0.5 MeV/A (open triangles) which in the same range that the
radial collective energy estimated in both QP sources
(full squares) and ISIS data (full triangles). This coherence indicates a
general property of nuclei : the relation between thermal pressure and
thermal energy independently of the formation process.\\
For sources produced in peripheral collisions, the friction/abrasion process
provides sources with thermal pressure only. Such sources, starting from
normal density, 
directly enter the low density region without an intermediate step in the
high density region as it is the case for central collisions.
These two types of sources provide a good opportunity to track possible influences of different trajectories
in the phase diagram on the fragment partitions.

\section{Comparison of fragment partitions.}
For a given excitation energy, the total charge bound in fragments ($Z_{frag}$) scales
with source charge ($Z_{s}$) for both QP and QF sources. To see how the
bound charge is shared among the fragments, we start
comparing the charge of the biggest fragment, $Z_{1}$, in each event. Many
thermostatistical studies \cite{iwm} indicate that
$Z_{1}$ is a good candidate as order parameter of phase transition in hot nuclei. We observe that
its mean value is governed by excitation energy and largely independent of
source sizes and production mode, which is not the case if we look to
$Z_{1}$ fluctuations. Indeed, on the excitation energy range studied, trends
are similar but $Z_{1}$ fluctuations for QP sources
 are larger and in good agreement with systematics reported in
\cite{wci}.\\
To go further, we propose to study the behaviour of the $A_{Z}$ observable
(eq.\ref{az}) that we name the generalized
asymmetry. It summarizes information on fragment charge partition
independently of the fragment multiplicities which largely differ for QP and QF sources.
The main result, comparing the behaviour of $A_{Z}$, is that
partitions from QP sources are more asymmetric
than those from QF sources. Since, the charge bound in fragments
($Z_{frag}$) scales with the size of the
source ($Z_{source}$) and the mean charge of the biggest fragment $Z_{1}$ is
quasi independent of the different sources, one
must consider a possible bias due to the specificity of the largest
fragment. To do that we can now take the fragment 
partition but without the biggest fragment in the calculation of the
generalized asymmetry that we call $A_{Z}\backslash\{Z_1\}$. The effect
still remains, which confirms the difference between QP and QF sources as
far as charge partitions are concerned.
\begin{center}
\begin{eqnarray}
A_{Z} = \frac{1}{\sqrt{M_{frag}-1}}\frac{\sigma_{Z}}{<Z>}\\ \label{az}
\Delta O = \frac{<O>^{Xe+Sn} - <O>^{QPAu}}{<O>^{QPAu}}\biggr|_{E^{*}} \label{diff}
\end{eqnarray}
\end{center}
\section*{Conclusions and outlooks.}
\begin{figure}
\begin{center}
\includegraphics[scale=0.45]{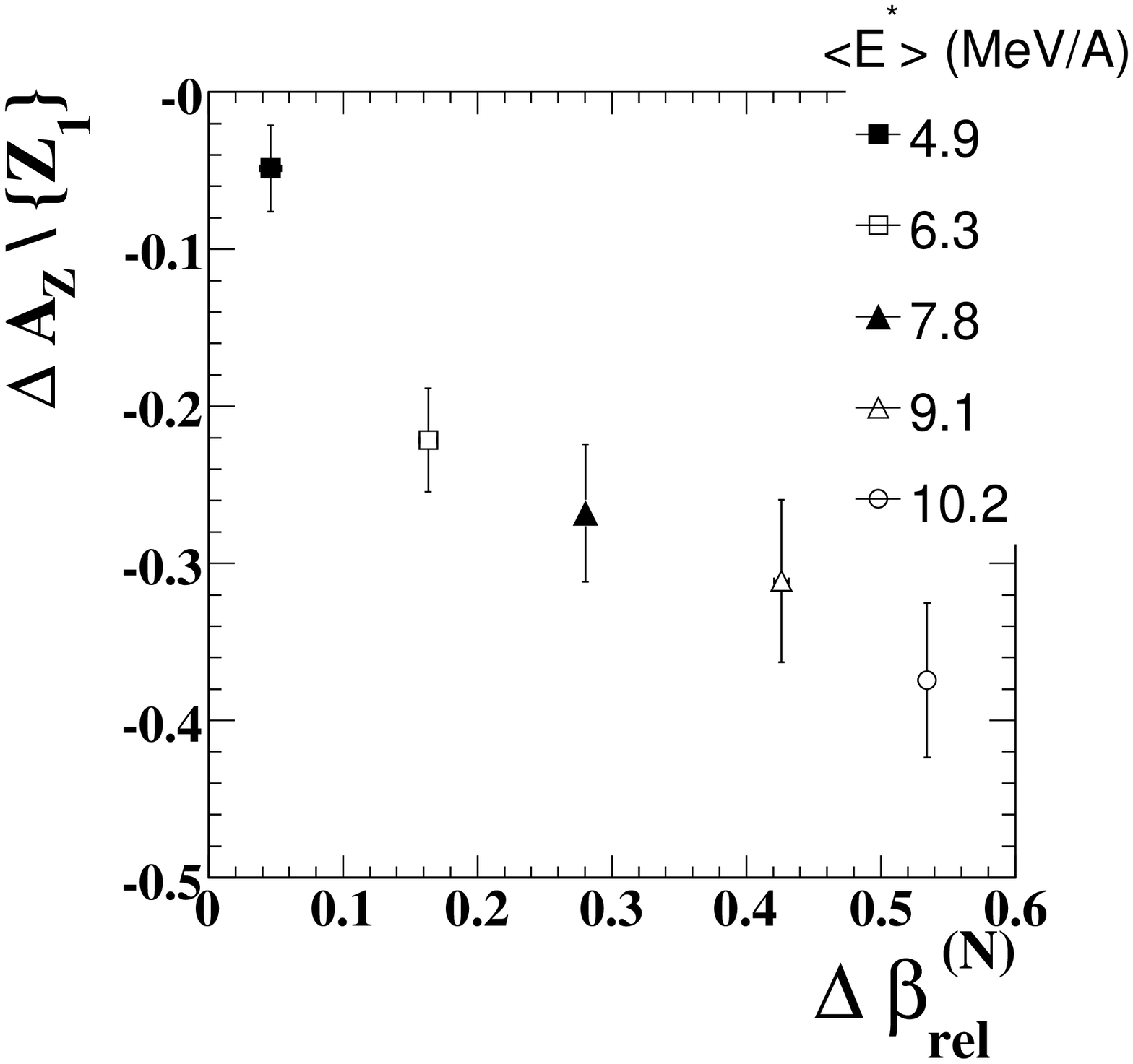}\caption{}
%\caption{Correlation between radial collective expansion and fragment partition asymmetry (see text)}
\label{fig02} % optional figure label, must be unique
\end{center}
\end{figure}
In this work, we have used two event ensembles provided by heavy-ion
collisions in the intermediate energy domain,
Quasi-Fusion and Quasi-Projectile sources, to deepen our knowledge of
multifragmentation. If the general behaviour of multifragmentation 
(total charge bound in fragments, charge of the biggest fragment in each event) is
governed by the excitation energy
deposited into the sources , differences at
a given excitation energy have been observed in both the kinematic
properties and partitions
of fragments for the two types of sources.
Using previous estimates of
radial collective energy in central
collisions, a calibration of the normalized relative velocity for fragments $\beta^{(N)}_{rel}$
was made. Extracted radial collective energy values for QP sources are found
much lower than those of QF sources and in good agreement with data from multifragmentation reactions induced by
hadrons. Such low values are compatible with radial collective energies only
produced by thermal pressure. Fragment
charge partitions of the two types of sources show differences in the fragmentation degree ($A_{Z}$ observable)
or in the charge fluctuations of the largest fragment ($\sigma_{Z_{1}}$).
To go further and link differences in trajectories in the phase diagram  and partitions, we propose
to study the correlation between dynamic and static observables. We define the relative
difference ($\Delta O$) between the mean values
of a given observable $O$ for QP and QF sources (eq.\ref{diff}). Taking as
reference the QP ones (radial collective thermal energy only), one can obtain an estimate of
compression/expansion effects for QF sources.
We calculate this
relative difference for $A_{Z}\backslash\{Z_1\}$ and $\beta^{(N)}_{rel}$ observables and for
the five common excitation energy of sources. 
As preliminary result, fig.\ref{fig02} shows a strong correlation which
indeed indicates that radial collective expansion is responsible for
more symetric fragment partitions in multifragmentation of QF sources.
%%%%%%%%%%%%%%%%%%%%%%%%%%%%%%%%%%%%%%%%%%%%%%%%%%%%%%%%%%%%%%%%%%%ù

%%%%%%%%%%%%%%%%%%%%%%%%%%%%%%%%%%%%%%%%%%%%%%%%%%%%%%%%
% End of the paper
%
\end{document}